# From High-Entropy Alloys to Alloys with High Entropy: A New Paradigm in Materials Science and Engineering for Advancing Sustainable Metallurgy

Jose Manuel Torralba [a, b,1,*], Alberto Meza [b,1], Venkatesh Sivagnana [a, b], Amir Mostafaei [c], Ahad Mohammadzadeh [d]

[a] Department of Materials Science and Engineering and Chemical Engineering. Universidad Carlos III de Madrid, Av. Universidad 30, 28911, Leganés, Spain

[b] Imdea Materials Institute, Calle Eric Kandel, 2, 28906, Getafe, Madrid, Spain

[c] Department of Mechanical, Materials, and Aerospace Engineering, Illinois Institute of Technology, 10 W 32nd Street, Chicago, IL, 60616, USA

[d] Department of Materials Engineering, Faculty of Engineering, University of Maragheh, Maragheh, P.O. Box 83111-55181, Iran

[1] equally contributed first authors

[*] Corresponding author: Jose Manuel Torralba (torralba@ing.uc3m.es)

## Highlights

- The transition from the concept of HEAs, also known as MPEAs, to AHEs is analyzed.

- High-entropy, stacking fault energy, lattice misfit and anti-phase boundary energy are introduced as design parameters for new alloys.

- Influence of these parameters in developing metallic advanced materials is discussed.

- A new paradigm in the design of new alloys based in a new way to correlate performance, microstructure, manufacturing and chemical composition is proposed.

## Abstract

The development of high-entropy alloys (HEAs) has marked a paradigm shift in alloy design, moving away from traditional methods that prioritize a dominant base metal enhanced by minor elements. HEAs instead incorporate multiple alloying elements with no single dominant component, broadening the scope of alloy design. This shift has led to the creation of diverse alloys with high entropy (AHEs) families, including high-entropy steels, superalloys, and intermetallics, each highlighting the need to consider additional factors such as stacking fault energy (SFE), lattice misfit, and anti-phase boundary energy (APBE) due to their significant influence on microstructure and performance. Leveraging multiple elements in alloying opens up promising possibilities for developing new alloys from multi-component scrap and electronic waste, reducing reliance on critical metals and emphasizing the need for advanced data generation techniques. With the vast possibilities offered by these multi-component feedstocks, modelling and Artificial Intelligence based tools are essential to efficiently explore and optimize new alloys, supporting sustainable progress in metallurgy. These advancements call for a reimagined alloy design framework, emphasizing robust data acquisition, alternative design parameters, and advanced computational tools over traditional composition-focused methodologies.

**Keywords**: high entropy steels; high entropy superalloys; high entropy intermetallics; sustainable metallurgy; new materials design paradigm.

1. **Introduction**

The emergence of high entropy alloys (HEAs), also known as multi-principal element alloys (MPEAs), represents a paradigm shift in alloy design and development [1]. Traditionally, alloys consisted of a base metal "enhanced" by adding minor elements that complemented its virtues or minimized its shortcomings. The number and quantity of alloying elements were typically limited by the potential formation of unwanted secondary phases, such as brittle intermetallic [2]. However, in 2004, both Cantor [3] and Yeh [4] proposed the possibility of creating monophasic microstructures from a mixture of five or more elements forming a single solid solution. This proposition fundamentally altered the criteria for achieving a solid solution, moving away from the Hume-Rothery rules [5]. The most important design characteristic became mixing entropy because achieving high mixing entropy in the designed alloy increases the Gibbs energy required to form intermetallic compounds and can promote the formation of a unique single phase [1].

MPEAs have demonstrated their ability to exhibit a wide variety of properties. Tsai [6] highlights their value in applications related to varying physical properties, such as electrical or thermal conductivity, and their suitability for extreme corrosion conditions [7–9]. MPEAs also show promise due to their excellent mechanical performance [10,11], even at high [1] or cryogenic temperatures [12]. Reviews on their mechanical behavior, fundamental deformation mechanisms [13] and fracture resistance [1] are available, as well as two noteworthy books on the subject [14,15]. Additionally, three reviews focused on powder metallurgy and HEAs have been published [16–18], including one on additive manufacturing (AM) [19].

In 2021, a report by 15 international experts, funded by two American defence agencies, was published by the Minerals, Metals & Materials society (TMS) [20], outlining the pathways to realizing the revolutionary potential of HEAs. The report offered nine recommendations, including "high-throughput screening methods and experimental tools," "predictive structure-property models and computational tools" "high-temperature testing and processing equipment," "in situ characterization methods," and "thermodynamic databases for complex concentrated alloy combinations (calculation of phase diagrams (CALPHAD))"[21]. Despite technological advancements, significant hurdles remain in the broader application of Artificial Intelligence to accelerate HEA discovery. The scarcity of labelled and unlabelled datasets (especially those correlating processing parameters, microstructural features, and resulting properties) impedes robust model training[22–30]. Moreover, existing machine learning (ML) approaches often lack critical manufacturability inputs, including thermal stability and oxidation behaviour, thus generating predictions that may be infeasible for real-world fabrication [31–37]. Incorporating thermodynamic simulations with ML-based modelling can help address this gap, supplying physics-informed insights into phase formation rules, melting behaviour, and diffusion kinetics [38–43].

Further development of HEAs represents a transformative shift in materials science, particularly in the pursuit of sustainable metallurgy. Sustainability in HEAs can be categorized into direct and indirect sustainability aspects. Direct sustainability is achieved through strategies such as utilizing low-carbon and energy-efficient feedstocks, shifting from equimolar to optimized alloy compositions, and promoting secondary and tertiary synthesis via recycling. Indirect sustainability, on the other hand, refers to HEAs' potential to enhance the efficiency and lifespan of downstream products, thus offsetting their high production footprint. For instance, certain HEAs demonstrate superior corrosion resistance, catalytic activity, hydrogen embrittlement resistance, and recyclability, making them viable candidates for high-efficiency energy conversion systems, lightweight structural applications, and biomedical devices[44]

The classification of HEAs has also evolved, extending beyond their initial definition based on equimolarity. Modern HEAs can be divided into FCC-, BCC-, and dual-phase HEAs, each offering distinct mechanical, thermal, and chemical properties. For example, FCC-based HEAs[45]are known for their high ductility and corrosion resistance, making them suitable for biomedical and cryogenic applications. In contrast, BCC-based HEAs, including refractory HEAs, exhibit high strength and oxidation resistance, positioning them as promising materials for aerospace and high-temperature applications. Moreover, dual-phase HEAs combine the advantages of both FCC and BCC structures, optimizing the trade-off between strength and ductility[44].

Recent studies on FeCoNiCuAl high-entropy alloy films (HEAFs) have highlighted their outstanding corrosion resistance, particularly after annealing, which enhances the stability of the passivation film and promotes the formation of high-valence oxides. This improvement results in superior electrochemical performance compared to 304 stainless steel[46]. Additionally, the TiNbTaZrMo HEA system has emerged as a promising biocompatible alloy, where recent porosity modeling studies have revealed that controlled porosity significantly impacts mechanical properties, such as Young's modulus, which can be fine-tuned to match human bone[47]. This makes HEAs highly relevant for orthopedic implants and dental applications, where an optimal balance between mechanical integrity and biological integration is essential.

While these new alloys have shown a remarkable capacity to surpass established limits in many applications, some challenges remain. Two primary concerns are the high demand for raw materials, which often include critical and strategic elements, and the limited market availability of ready-to-process alloys [48]. Initially, this new approach to alloy design focused on equiatomic mixtures of five or more elements (HEAs) [1]. The concept has since expanded to include non-equiatomic alloys (MPEAs), where no single component dominates, as is common in classical metallurgy [49]. Recently, it has even been observed that certain properties in HEAs can be further improved with the addition of minor elements [50,51].

The exceptional properties achieved with HEAs are attributed to four core effects [52–55]: the cocktail effect, sluggish diffusion, severe lattice distortion, and high entropy of mixing. Recent studies, however, suggest a fifth core effect—the presence of short-range orderings, which can significantly contribute to strengthening [56–58].

The emergence of HEAs/MPEAs revolutionized alloy design, but these alloys came with certain limitations. Initially, compositions were restricted to equiatomic ratios, though this requirement was later reduced [59]. Other compositional restrictions remain, including limitations on minority elements that can affect microstructure design and the absence of a 'majority component' (e.g., Fe, Ni, or Co).

This viewpoint paper proposes a radical shift in alloy design to overcome these restrictions. By designing microstructures based on the high configurational entropy of the mixture, we remove constraints on alloying elements. Both majority and minority elements can be used freely, enabling any alloying element to contribute positively to the design without compositional restrictions. This approach moves beyond HEAs/MPEAs and rather towards designing alloys with high entropy (AHEs) such as high entropy steels, high entropy superalloys, or high entropy intermetallics.

## 2. High entropy steels

In 2015, Raabe [60] introduced the concept of "high entropy steels" based on the idea that entropy-driven stabilization of massive solid solutions can be achieved even in non-equiatomic quinary (or higher) alloys. This approach broadens the use of minor elements, which can contribute to increasing configurational entropy, thus maximizing the likelihood of forming a solid solution. However, it is also suggested that maximizing entropy does not necessarily minimize the formation of intermetallic phases. Nevertheless, in some steels, this intermetallic formation can actually be advantageous, providing additional strengthening through solution hardening and aging processes.

Raabe's work also emphasizes that high entropy alloys can be successfully developed without adhering to equiatomic compositions. A particularly valuable contribution from this work is the introduction of SFE as a design parameter. Low SFE can be associated with the TRIP/TWIP effect (transformation induced plasticity (TRIP) and twinning induced plasticity (TWIP)), leading to significant improvements in mechanical properties such as ductility and toughness[61,62]. These effects were also noticed in HEAs [63]. Steel compositions can be tailored to achieve this low SFE, with elements such as carbon or manganese playing key roles in fine-tuning it [60].

Figure 1 summarizes the tensile properties from various high-strength steels. The mechanical properties of the developed high entropy steels [60] surpass those of many conventional austenitic Fe–Cr–Ni steels, with ultimate tensile strengths reaching up to 1000 MPa and tensile elongations extending to 100% in some cases. This newly developed philosophy in steel design opens the door to pushing the mechanical properties of steels far beyond the current state of the art.

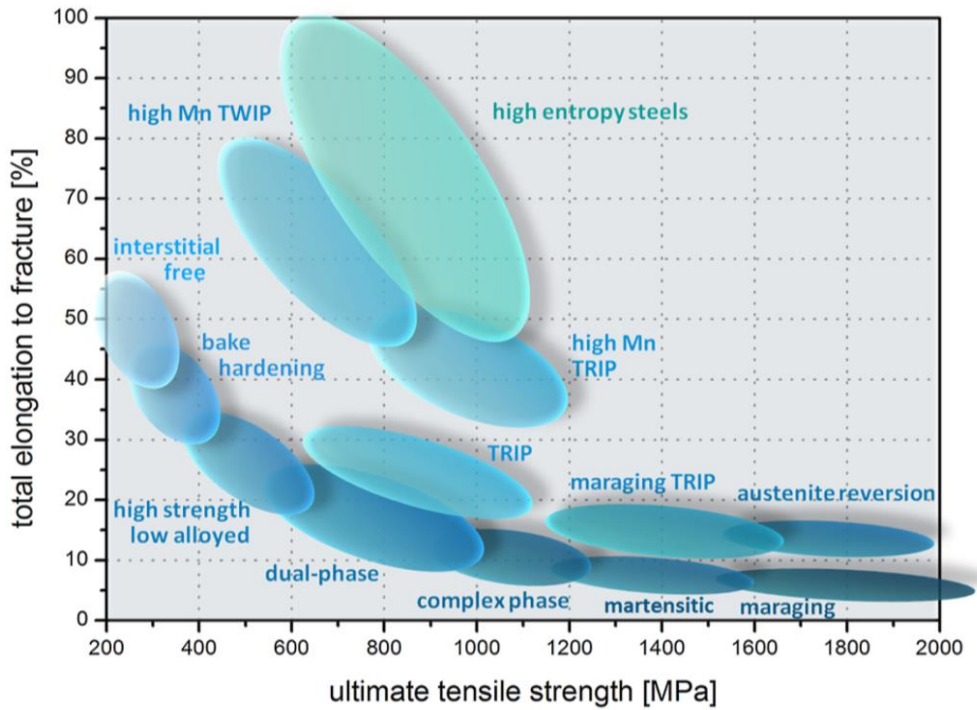

Figure 1. Tensile properties of the high entropy steels in comparison with many conventional austenitic Fe–Cr–Ni steels [60].

Numerous studies applied the concept of high entropy alloy design to the development of Fe-based high entropy alloys, commonly referred to as high entropy steels [64–66]. In several of these studies, the low SFE design criterion has been introduced. In single-phase FCC alloys, such as many high entropy alloys including steels, the TRIP/TWIP effect can occur, leading to a significant improvement in mechanical properties. As mentioned earlier, this TRIP/TWIP effect is often associated with low SFE [67,68]. Moreover, a correlation has been observed between low SFE and superior cryogenic behavior [65,67–69], as well as excellent tensile properties coupled with high ductility [70]. SFE has also been linked to the increasing number of alloying elements [71] and the rise in the percentage of γ-stabilizers in FCC steels [72–74]. Therefore, low SFE should be considered a key selection criterion, alongside high configurational entropy of mixing, in the development of high entropy alloys.

These concepts together offer the potential for developing high entropy steels that can be strengthened either through precipitation or the TRIP/TWIP effect [75,76]. By incorporating both high configurational entropy and low SFE as design criteria, it becomes possible to utilize multiple elements, avoiding the exclusion of certain elements commonly found in standardized steels.

To further expand the scope of high entropy Fe-based alloys, high entropy white cast iron has also been developed using this approach [77–79]. In this case, the high entropy effect is not utilized in the microstructure. Instead, following the MPEA concept, a large number of carbide-forming elements are added to a cast iron base composition (e.g., Fe-20Cr-5C). In the molten state, configurational entropy is high, but during solidification, these carbide-forming elements compete to form carbides. This competition suppresses carbide growth, leading to refined carbides. The primary goal here is to increase hardness, which is successfully achieved as the microstructure retains the typical pattern of white cast iron but with a significant increase in fine carbide precipitates [78]. Figure 2 shows how increasing the amount of competing alloying elements to produce uniformly distributed fine carbides. In Alloy #1 it can be seen being a hypoeutectic cast iron with primary austenite and eutectic carbides. The other four alloys are designed near the eutectic point to improve hardness without coarse primary carbides. We can see in alloys #2-5 how adding B, vanadium, and niobium the microstructure is refined. The higher magnification reveals phases such as primary austenite, $M_7C_3$ carbides, and vanadium and niobium carbides. Alloy #4 shows a eutectic-cell structure with tungsten carbide, and Alloy #5 has a much complex microstructure.

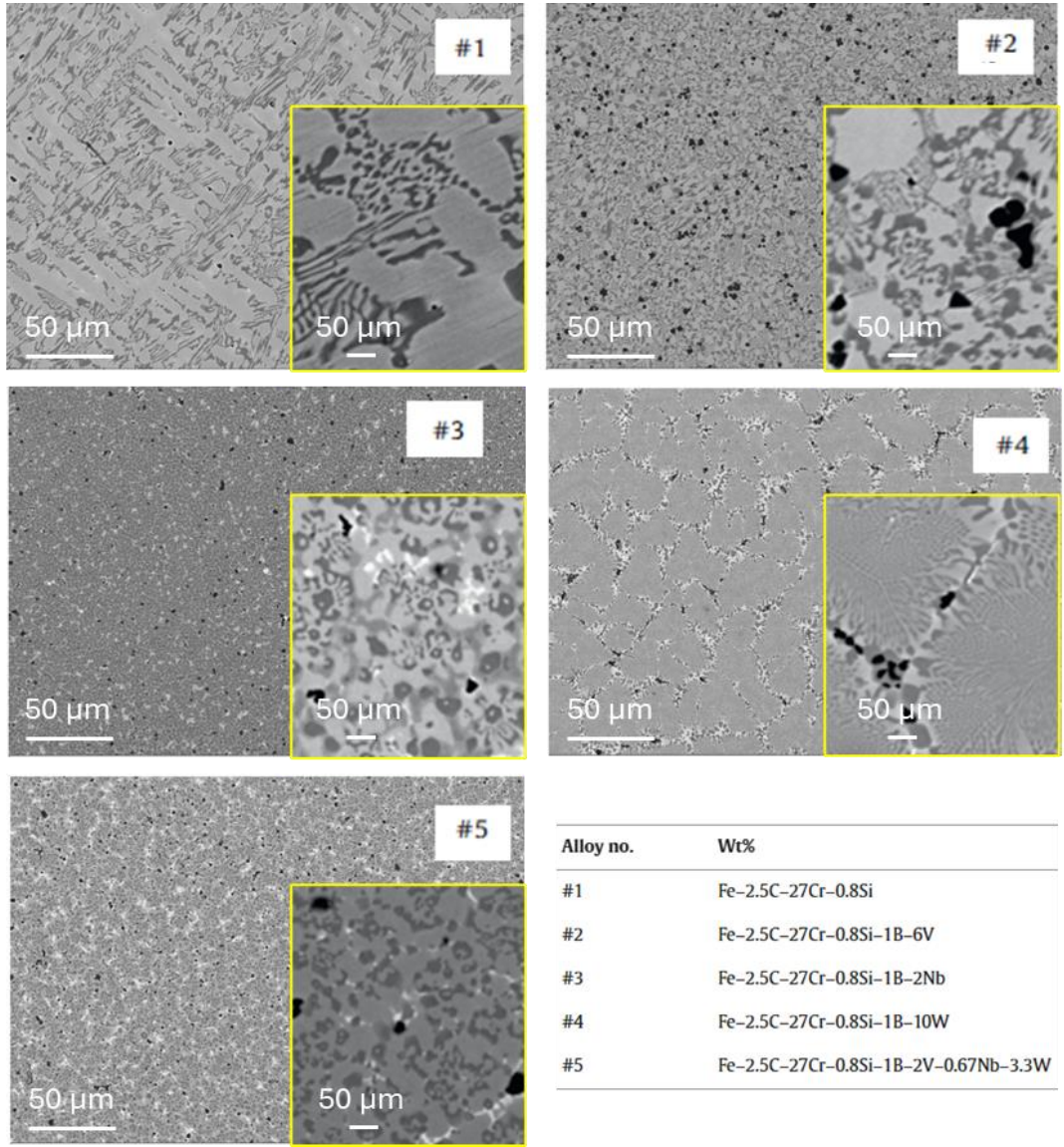

Figure 2. Microstructure of 5 fabricated cast iron with high entropy [78].

## 3. High entropy superalloys

The concept of high entropy superalloys (HESA) was introduced around the same time as other high entropy steels [80–82]. However, two distinct families of HESAs have emerged, each following a different approach to microstructural evolution. The first family mirrors the strategy used in high entropy steels, where a high entropy of mixing is achieved by accumulating multiple alloying elements to form a single solid solution. Research in this area has focused on refractory elements, leading to the development of a

BCC+B2 microstructure, where a disordered BCC phase serves as the matrix for an ordered BCC (B2) phase that appears as cuboids [80,82,83]. This microstructure is analogous to the traditional γ-γ' structure seen in superalloys. In Figure 3, the FCC/B2-BCC microstructure of a high-entropy alloy with the composition Ni19.5Fe19.3Cr21.1Co19.6Al20.5 is clearly visible. A dual-phase microstructure can be identified, where the FCC phase surrounds the BCC-rich regions. In Figure 3(b), the typical spinodally decomposed structure is evident within the BCC phases. This spinodal decomposition plays a crucial role in enhancing the alloy's strength due to the coherent interaction between the two BCC phases and the formation of fine-scale modulated structures. Additionally, the morphology of the inner part of the FCC network closely resembles a γ-γ' structure, where the B2 phase acts as the matrix, analogous to the γ matrix in conventional superalloys.

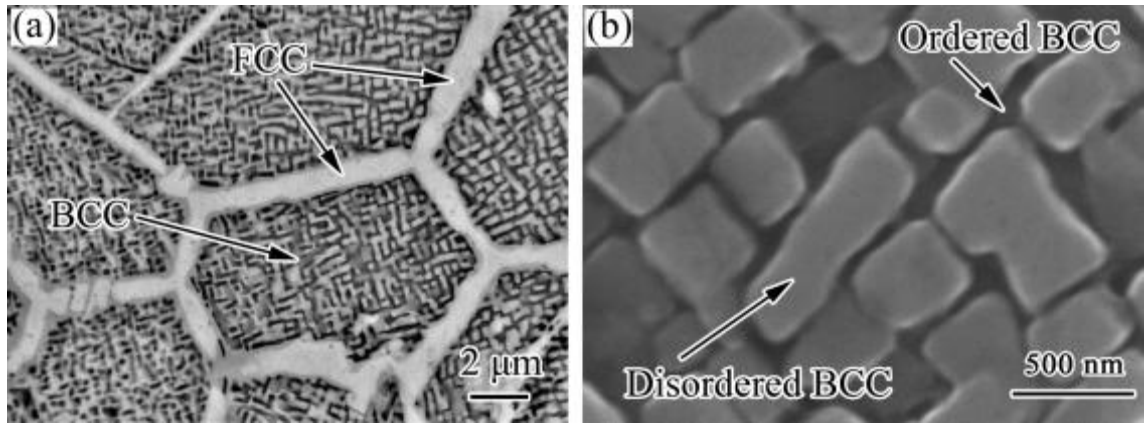

Figure 3. Microstructures of the spark plasma sintered AlCoCrFeNi HEA: (a) low magnified back-scattered electron image, (b) high magnified secondary electron image of the disordered BCC phase and the ordered BCC phase (B2)[84].

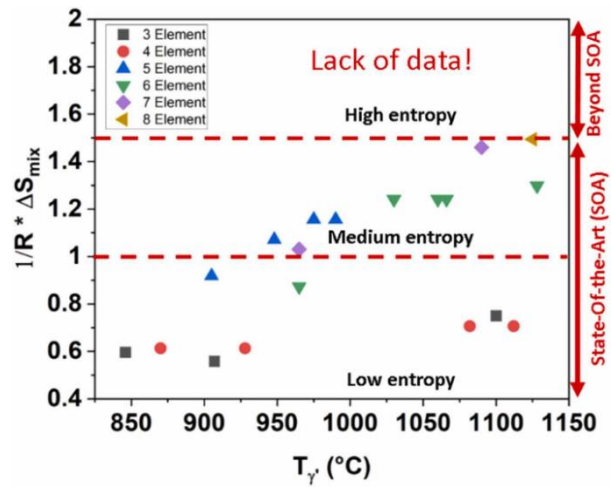

Figure. 4. Relationship between $\Delta S_{mix}$ and γ' solvus temperature in Co- or CoNi-based superalloys[85], data was obtained from many sources in the literature [86].

The second family of HESAs is based on adding multiple elements to a Ni or CoNi base alloy to ensure high mixing entropy [81,87–90]. This approach typically results in a γ-γ' microstructure, where the γ matrix has a high mixing entropy. Some studies have shown that the high mixing entropy character of these superalloys raises the solvus temperature, thereby enhancing their stability at high temperatures [81,86,87] (as can be seen in Figure 4). Based on the historical development of superalloys and a comprehensive analysis of data published in the literature, Mohammadzadeh et al. [86] has been identified a relationship between the entropy of mixing ($\Delta S_{mix}$) and the γ' solvus temperature in Co- or CoNi-based superalloys, as illustrated in Figure 4. For instance, enhancing $\Delta S_{mix}$ from 0.613R (four elements) to 1.298R (six elements) raises the γ' solvus temperature from 870 °C to 1128 °C[91]. The figure suggests that designing novel alloys based on the high entropy concept[52], i.e., HEAs, holds promise in overcoming the limitations of the γ' solvus temperature in Co-based superalloy applications. However, it is worth noting that while researchers may be exploring the addition of further elements in Co- or CoNi-based superalloys, they may not have investigated the role of $\Delta S_{mix}$ or the principles of HEAs in their design strategies ("lack of data" as shown in Figure 4). So, it seems that further development of Co- or CoNi-based superalloys must consider the HEAs principles as an essential factor.

The development of HESAs based on refractory metals such as Mo, Nb, Ta, and Zr, along with Al and Ti, was introduced by Senkov et al [80]. The microstructural evolution of these alloys is driven by spinodal decomposition. At high temperatures, a BCC phase is stable, but upon cooling, a dual microstructure forms, consisting of an ordered cubic B2 phase, rich in Al (plus Ti and Zr), within a disordered BCC phase, rich in Ta (plus Nb and Mo), with a cuboidal shape. Assuming that a single BCC or B2 phase-field exists at high temperature, Dasari et al. [92] illustrates the phase separation of the single BCC or B2 phase into continuous BCC matrix and discrete B2 precipitates in three different stages in Figure 5. In stage 1, the single BCC or B2 phase decomposes to produce a nearly co-continuous mixture of BCC and B2 phases which could have formed by spinodal decomposition followed by ordering or disordering. In stage 2, the co-continuous mixture evolves into a microstructure with continuous B2 phase and discrete BCC precipitates. Stage 3 involves phase inversion to produce a microstructure consisting of a continuous BCC phase with discrete B2 precipitates.

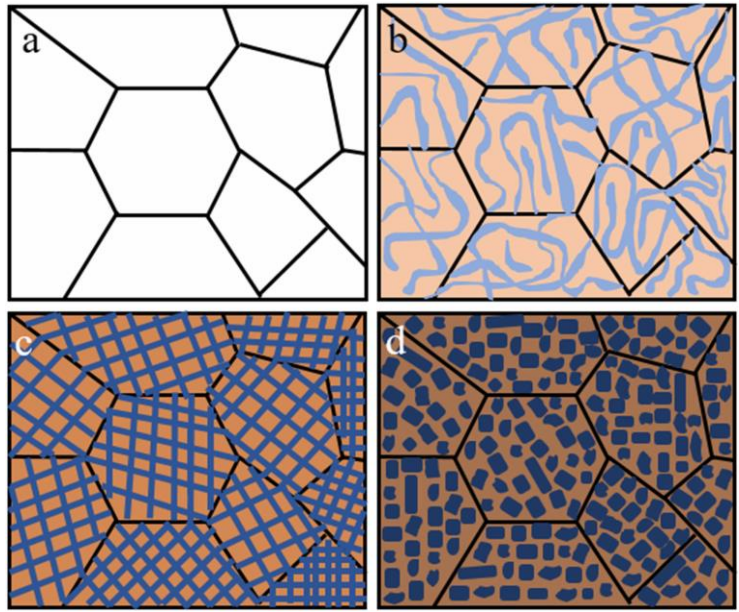

Figure 5. (a) A single-phase BCC or B2 microstructure existing before the transformation, (b) Stage 1: A co-continuous BCC (orange) + B2 (blue) microstructure, (c) Stage 2: The checkered pattern microstructure at an intermediate state where B2 is continuous, and BCC is discrete, (d) Stage 3:Near-equilibrium microstructure with continuous BCC and discrete B2 phases [92].

Both phases share the same crystal orientation. As has been mentioned, this microstructure is similar to the γ-γ' structure of Ni-based superalloys, enabling these HESAs to perform exceptionally well at high temperatures. Figure 6 compares the yield strength and specific yield strength of these HESAs with some nickel-based superalloys [80]. These HESAs can be classified as refractory high entropy superalloys (RHEA).

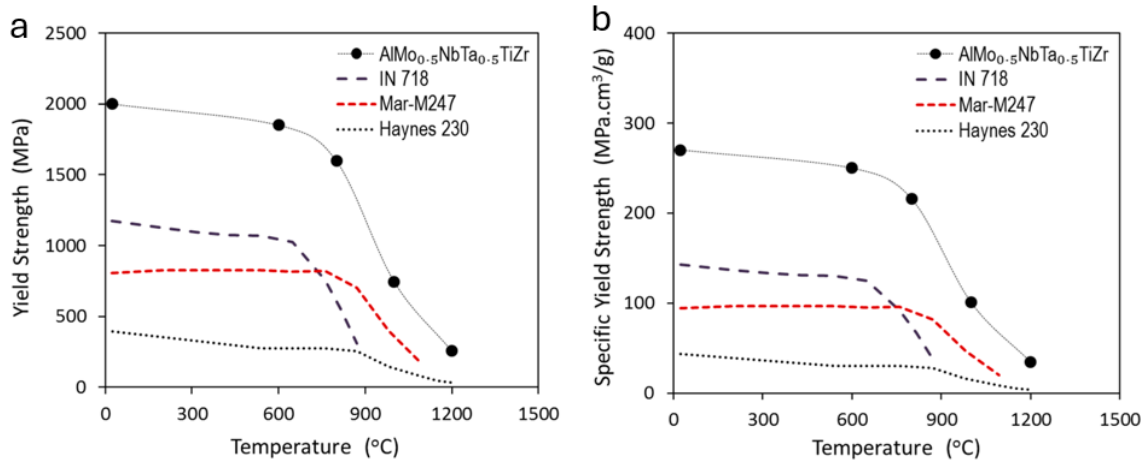

Figure 6. Comparison of the strength of the RHEA superalloy AlMo0.5NbTa0.5TiZr and three Ni-based superalloys obtained by precipitation hardening IN718, Mar-M247, and Haynes® 230 [80].

It is noteworthy that despite their high-density refractory metal components, this family of alloys can compete favourably in specific properties (normalized by the density). Miracle et al. [93] provided an insightful short review of this new family of RHEAs. Over time, this family of RHEAs has evolved to incorporate additional strengthening mechanisms, including the precipitation of intermetallic phases (hexagonal or orthorhombic), which help to increase hardness and compressive strength [83,94–98]. The microstructural effect of these precipitates after a solution and aging treatment in an RHEA [99] can be seen in Figure 7. As previously stated, the primary phases in RHEAs are a disordered BCC phase and an ordered BCC phase (B2 phase), (Figure 3). Since 2022, some authors [100–102] have begun referring to the disordered BCC phase as the A2 phase, following the "Strukturbericht" designation [103], to avoid misunderstandings.

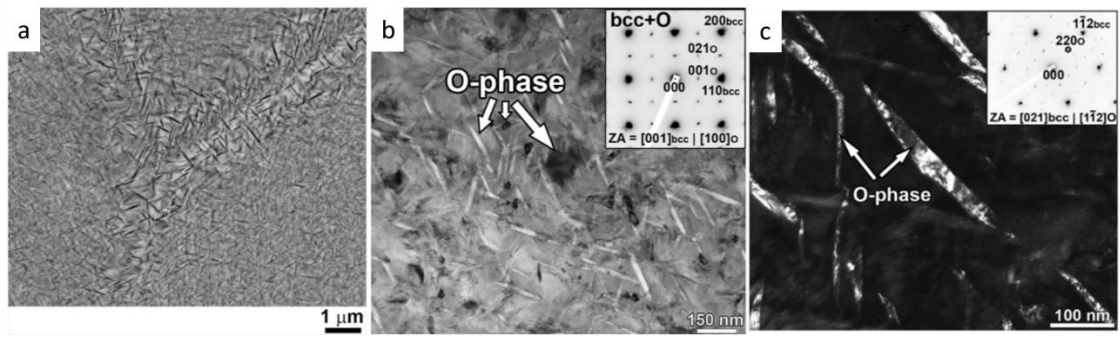

Figure 7. Microstructure of the $Ti_{40}Nb_{30}Hf_{15}Al_{15}$ alloy after annealing at 1200 °C and further aging at 600 °C (c-e); (a) SEM-BSE image; (b) TEM brightfield images; (c) TEM dark-field image taken from a diffraction spot (220) in [112] zone axis of the orthorombic-phase (corresponding selected area electron diffraction (SAED), pattern is shown in the inset) [99].

The second strategy for developing HESAs is based on utilizing the same strengthening mechanisms as conventional Ni-based superalloys, specifically the development of a disordered FCC matrix (γ) reinforced by ordered FCC intermetallic precipitates (γ'). Yeh et al. [81] introduced this family of HESAs in 2015. The goal was to achieve a stable FCC γ matrix with high entropy of mixing by incorporating a large number of alloying elements, using Ni as the principal component. By using a diverse array of alloying elements, the entropy level can be optimized, and specific final properties can be tailored. The study focused on three different alloys within the Ni-Al-Co-Cr-Fe-Ti system. The development process utilized the CALPHAD method to avoid the formation of undesirable intermetallic phases, such as the Laves phase. In Figure 8 [81], two phase diagrams of those studies using CALPHAD to avoid undesired phases such as topologically closed packed (TCP) phases are shown. In these base compositions, transition metals were used, and the use of high-density refractory metals was minimized (with only small quantities of Ta, Mo, and W), ensuring densities below 8 g/cm³. The presence of Al and Cr also provides good oxidation resistance at high temperatures. The strengthening mechanism in these HESAs is γ' intermetallic precipitation through solution/aging treatment. The large number of alloying elements also ensures the proper misfit between the γ and γ' phases leading to improved creep resistance.

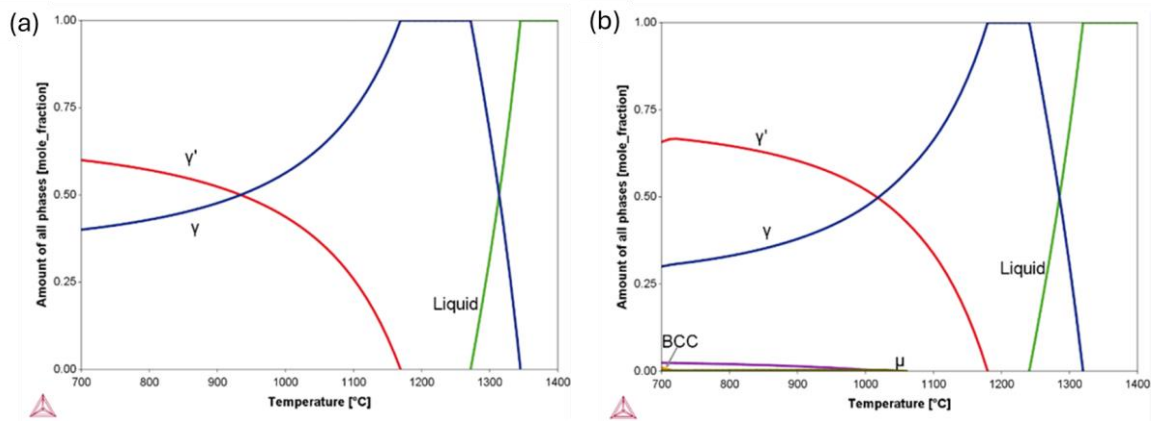

Figure 8: Designing alloy compositions by CALPHAD-base simulations (Thermo-Calc: TCNI5 database) (a) HESA $Ni_{50.5}Al_{8.9}Co_{9.2}Cr_{9.2}Fe_{8.2}Ti_{6.0}$, and (b) HESA $Ni_{48.6}Al_{10.3}Co_{17.0}Cr_{7.5}Fe_{9.0}Ti_{5.8}Ta_{0.6}Mo_{0.8}W_{0.4}$ [81].

As was mentioned above, according to various authors [81,86,87], the high mixing entropy level of these superalloys increases the solvus temperature of the γ' phase, enhancing performance at high temperatures. Tsao et al. [104] introduced two important design criteria for developing these HESAs including (1) proper lattice misfit between the γ and γ' phases and (2) high APBE. Lattice misfit (δ) is the difference in lattice parameters between γ and γ' phases. It was initially defined as "lattice mismatch" and can be calculated using the following expression as shown in Eq. (1) [105]:

$$\delta = \frac{2(a_{\gamma'} - a_\gamma)}{a_{\gamma'} + a_\gamma} \qquad \text{Eq. (1)}$$

where $a_\gamma$ and $a_{\gamma'}$ are the lattice parameters of the γ and γ' phases, respectively. The misfit is strongly correlated with the morphology of the γ' precipitates. An example is shown in Figure 9 about how the misfit can influence the shape of the precipitates for different Ni base superalloys [106]. Figure 9 shows the evolution of the misfit with the square morphology. From sample DX-5 to D8 it increases with the square morphology and in the last sample DX-2 it decreases slightly due to the loss of quadrature. For low misfit values, the precipitates are spherical, and as the misfit increases, the shape tends to become cuboidal (initially weakly aligned, later perfectly aligned) [107,108]. This misfit is temperature-dependent, and at higher temperatures, the shape can change into elongated laths. High-temperature strength is correlated with a high volumetric fraction of ordered precipitates that resist shearing stresses [109]. It has also been demonstrated that creep resistance decreases with lower misfit values [110].

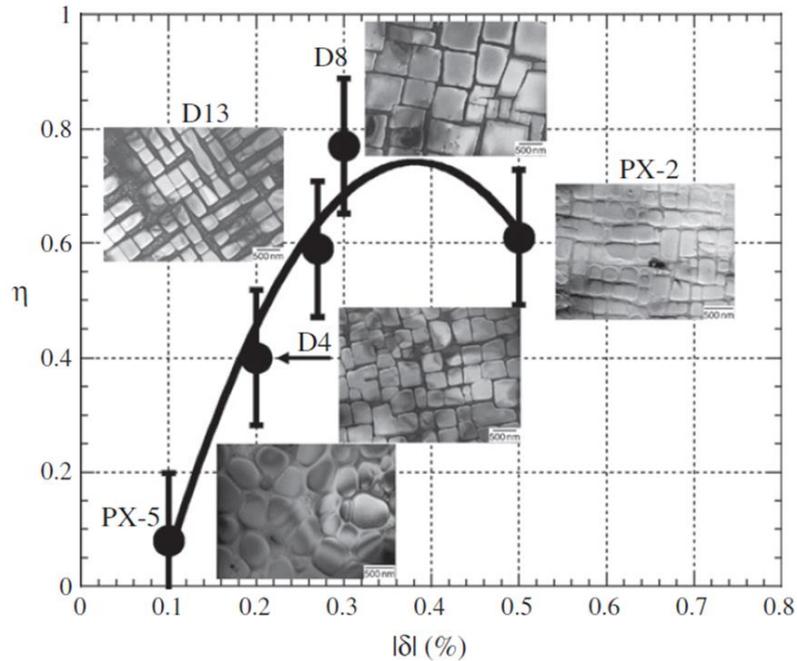

Figure 9. Shape parameter ratio as a function of high temperature lattice misfit magnitude. Compositions of plotted alloys are D4 ($Ni_{89.5}Ir_{2.5}Cr_5Re_1Ta_2$), D8 ($Ni_{89.5}Pt_{2.5}Cr_5Re_1Ta_2$), D13 ($Ni_{89}Pt_{2.5}Ir_{2.5}Cr_5Re_1$), PX-2 ($Ni_{75.9}Al_{14.5}Cr_{7.5}Si_{0.5}Mo_1Ti_{0.5}Hf_{0.1}$), and Px-5 ($Ni_{76.699}Al_{13.5}Cr_{7.5}Si_{0.5}Mo_1Ti_{0.5}Hf_{0.1}C_{0.1}B_{0.087}Zr_{0.014}$) [106].

In RHEAs, the misfit value can be determined by comparing the B2 and A2 phases, with high misfit values (associated with B2 cuboidal precipitates) shown to have a positive influence [94]. This beneficial effect of high misfit values on high-temperature behavior and creep resistance has also been observed in HESAs [81,87,111–114].

An APB is a planar defect that occurs within an ordered alloy structure, such as the γ' phase of nickel-based superalloys (see Figure 10). The γ' phase typically exhibits an ordered $L1_2$ crystal structure, which can be disrupted by dislocations moving through the precipitate, creating an APBE [115]. Higher APBE values indicate greater stability of the ordered structure, requiring more energy to create such defects. APBE plays a crucial role in the strengthening of superalloys, where the hardening mechanism is significantly influenced by resistance to dislocation movement [116,117]. Dislocations that traverse the γ' phase tend to form APBs, which enhance the alloy's strength by increasing the energy required for dislocation movement through the material. This effect also contributes to improved fatigue and creep resistance. Some authors consider APBE to be a primary factor in the strengthening of superalloys [112].

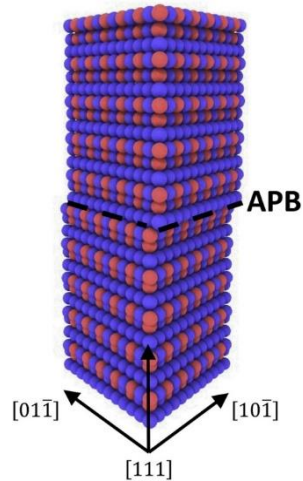

Figure 10. A typical atomistic structure used to calculate APBE. The APB was created by shifting the top crystallite with respect to the lower crystallite [118].

The APBE can be influenced by temperature and the composition of the γ' phase. An increase in temperature generally leads to a decrease in APBE, while elements such as Ti can increase it, whereas Cr and Co generally have a negative effect [71]. Reference [71] also describes various methods for calculating the APBE, most of which involve comparing the energy difference between an atomic volume with and without the APB. Numerous studies emphasize the critical role of APBE in the high-temperature properties of HESAs. For instance, Tsao et al. [87] highlight its influence on creep behavior. However, as mentioned earlier, the primary contribution of APBE is to shearing strengthening rather than precipitation strengthening [112]. In RHEAs, the role of APBE is also considered in alloy design, although a low APBE is sometimes desired to ensure greater ductility [93,102]. Joele and Matizamhuka [119] provide an extensive review of the strengthening mechanisms in HESAs.

## 4. High Entropy Intermetallic Compounds (HEICs)

The term HEICs was first introduced by Tsai [120] in 2016, although a paper published a year earlier had already reported the development of 'high entropy' intermetallics [121]. Intermetallics are a unique class of materials that, while composed of metals, exhibit properties more akin to ceramics. Although some intermetallics have variable stoichiometry, most possess fixed stoichiometry, similar to ceramics. They form not through solid solution but by exceeding the solubility limits of one component within another. They are a hybrid of metallic and ionic bonding, as seen in ceramics, and their crystalline structure is a superlattice with a high occupancy of interstitial voids, resembling ceramics. This combination of characteristics endows intermetallics with some advantageous properties of ceramics, such as excellent high-temperature performance, but also some drawbacks, such as limited ductility.

The crystalline structure of intermetallic consists of superlattices that can be decomposed into two overlapping simple lattices (Figure 11) [122]. In conventional intermetallic, one of these lattices is typically dominated by a single component, while the second lattice, usually interstitial to the first, is governed by one or two components. Table 1 lists the most important structures involved in the formation of intermetallic, according to their "Strukturbericht" designation [103].

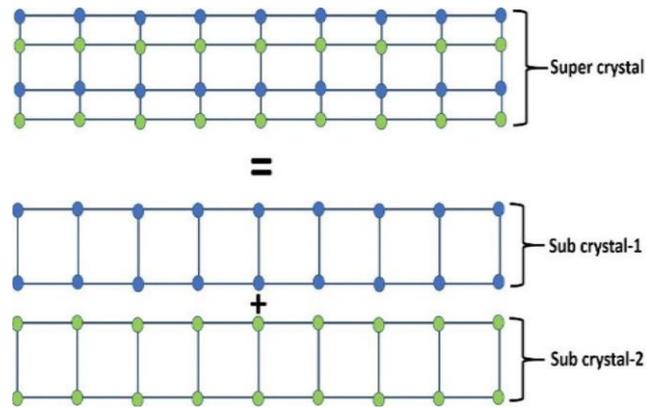

Figure 11. Schematic representation on the concept of formation of sub crystal and super-crystal [122].

In HEICs, sublattice 1 is typically occupied by one or two components, while sublattice 2 contains five or more components. Figure 5 shows a comparison between the superlattices of conventional intermetallics and those of HEICs across four different structures [123]. Figure 12 presents examples of developed HEICs.

Table 1. Most important structures in the formation of an intermetallic according to their "Strukturbericht" designation [103].

| Strukturbericht designation | Description |
| --- | --- |
| A1 | FCC crystals |
| A2 | BCC crystals |
| A3 | HCP crystals |
| B2 | Derivative of BCC in which lattice points are occupied by one atom and the body center point by another atom. |
| C14 | $MgZn_2$ laves phase |
| C40 | Silicide |
| $D0_3$ | A3B stoichiometry |
| $D0_{11}$ | Fe3C structure |
| $D0_{22}$ | Tetragonal |
| $D8_b$ | Sigma phase |
| $L1_2$ | Derivative of FCC crystal in which lattice points are occupied by one atom and the face center points are occupied by another atom. |
| $L2_1$ | Heusler alloy, prototype $Cu_2AlMn$ |
| S14 | Garnet structure |

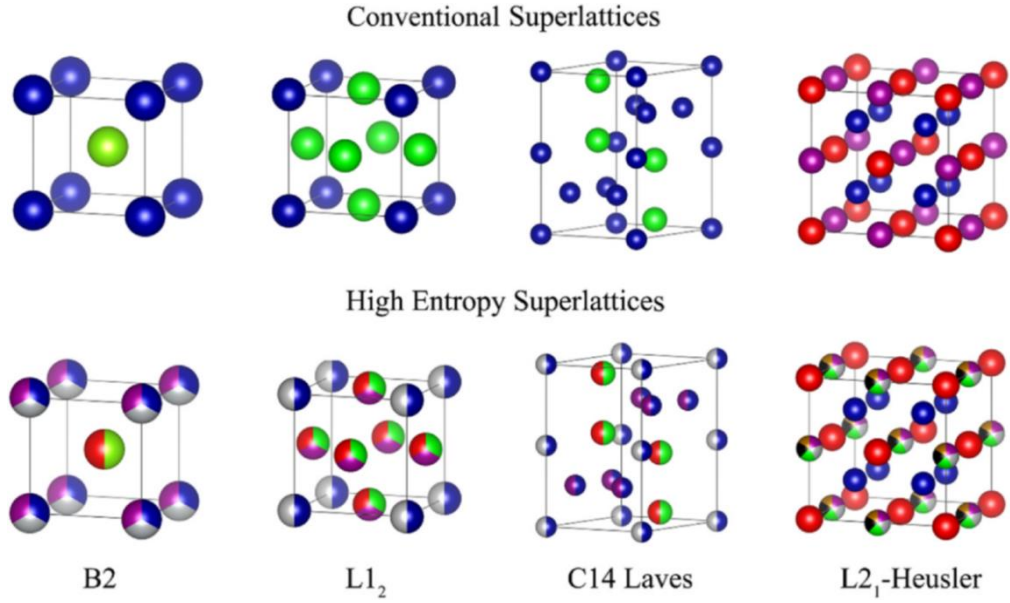

Figure 12. Schematic illustration of the atomic site occupation in conventional and high entropy B2, L12, C14 Laves, and L21-Heusler superlattices. The site occupancies shown for sublattices of HEICs are hypothetical and just indicate that each sublattice may be occupied by multiple elements [123].

According to Westbrook [124], the formation of conventional intermetallic can be attributed to two types of rules: electronic (such as valence electron concentration (VEC) and electronegativity) and geometric (like atomic size difference and packing efficiency). The VEC is also considered in the design of HEAs, while electronegativity, atomic size, and packing efficiency are related to Hume-Rothery rules. In the case of HEICs, additional design parameters must be introduced, including differences in atomic radii ($\delta_r$), electronegativity ($\chi$), average VEC, enthalpy of mixing ($\Delta H_{mix}$), ideal mixing entropy ($S_{id}$), and melting temperature ($T_m$). These terms, among others, are widely explained in the literature [1,125]. The method to calculate these parameters in materials with superlattices as their crystalline structure differs slightly, as Miracle and Senkov [1] suggest considering the influence of each parameter on each lattice independently for this family of materials. As an example, Eq. (2) shows how to calculate the configurational entropy of mixing for a HEIC [126]:

$$S^{HEI,ideal} = -k_B \left(\sum_{x=1}^{x} a^x \sum_{i=1}^{N} f_i^x \ln(f_i^x)\right) \bigg/ \sum_{x=1}^{x} a^x \qquad \text{Eq. (2)}$$

where $a^x$ is the number of sites on the sublattice x; $f_i^x$ is the fraction of element i randomly distributed on the sublattice x and N is the number of component elements. The methodology to calculate all the high-entropy parameters adapted to the specific characteristics of HEIC superlattices can be found in the work of Liu and Zhang [126].

HEICs find applications in various sectors. Moghaddam et al.[123] and Wang et al. [127] classify the good performance of HEICs into three domains: mechanical properties [126,128,129] (such as high-temperature performance and high compression yield strength), functional properties (including shape memory behavior [121], superconductivity, and magnetism [130]), and catalytic activity. This performance is notable even in comparison with conventional HEAs.

## 5. The development of sustainable alloys with high entropy

Improving alloy properties by adding more alloying elements has been a common approach in metallurgy. However, when performance limits are reached, physical metallurgy often steps in with heat and thermo-mechanical treatments, alongside alternative manufacturing methods. Consequently, high-performance alloys typically share several features including a complex and precise composition, restrictive manufacturing methods, and intricate treatments. Any deviation in composition or processing conditions often results in failure. This also means that the raw materials used are from a selected group of metals, many of which are considered critical.

The design of high-performance materials has been radically transformed by the emergence of HEAs or MPEAs and, more recently, by the concept of AHEs [131]. This approach shifts the design paradigm by using many alloying elements to achieve high configurational entropy. It eliminates previously restricted elements and minimizes the use of certain critical elements historically employed to achieve specific properties. Instead, the focus is on achieving high entropy and optimizing other design parameters such as lattice misfit, SFE, or APBE [13].

This new concept offers many possibilities for recycling and using waste to produce advanced materials. By leveraging high entropy, it is possible to create mixtures from standardized commercial alloys that meet the criteria for developing viable microstructures for high performance [132]. The advantages of this methodology include (1) the absence of prohibited elements (such as copper in most steels [133]), as all can contribute to increased configurational entropy; (2) the ability to avoid critical metals [134] as direct raw materials, since suitable mixtures can be achieved from commercial alloy scrap; and (3) the potential for direct use of multicomponent alloys from recycling domains such as mining residues and electronic scrap [48]. Furthermore, this approach promotes increased recycling of scrap (any scrap is useful), reducing energy and raw material consumption, and enhancing the recycling of electronic waste, which is currently minimal. Several studies [132,135,136] have demonstrated this approach by producing HEAs from blends of commercial commodity alloys.

Applying this philosophy to electronic waste (e-waste) could significantly improve recycling efficiency and facilitate the development of high-performance alloys without relying on critical metals as direct raw materials. The recycling of e-waste as a complex alloy simplifies the recycling process [137]. Most e-waste multicomponent alloys can be classified as nickel-based [138–140], copper-aluminium-iron-based [141–147], among others. Proper combinations of commodity alloys and e-waste can produce AHEs.

The feasibility of producing AHEs from e-waste has also been demonstrated [148]. Figure 13 shows (left) the Thermo-Calc (CALPHAD) phase prediction for a mixture obtained from e-waste compositions like smartphones, laptop printed circuits, and Li-ion batteries [143,149]. Figure 13 (right) displays the electron backscatter diffraction (EBSD) of this designed alloy obtained experimentally by arc melting, with the phases matching the predictions. These studies open the path toward the sustainable design of high-performance alloys from scrap and wastes. The next step is to define the methodologies required to achieve this goal.

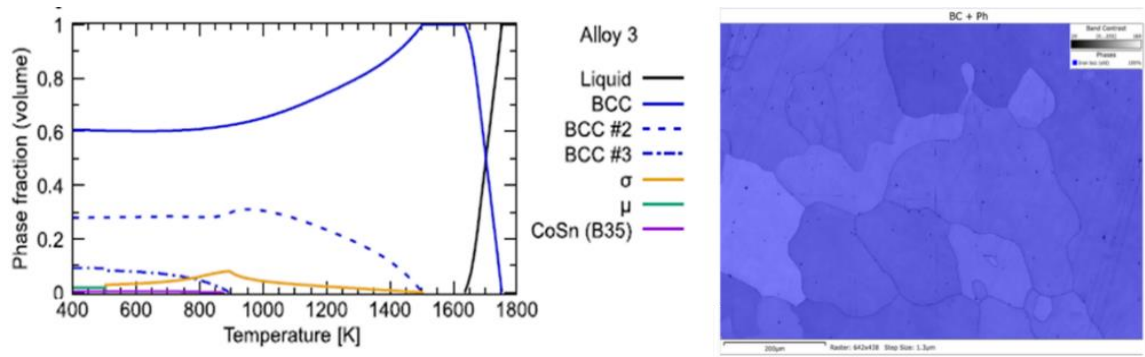

Figure 13. Left: Thermocalc phase prediction for a selected e-waste multiprincipal alloys (Al$_{29.1}$Fe$_{28.8}$Ni$_{19.3}$Co$_{11.8}$Cr$_{9.5}$Mn$_{0.6}$Sn$_{0.2}$); right: EBSD map of phases (BCC) [148]

## 6. Towards a new way of designing alloys: shifting the classical paradigm of Materials Science and Engineering (MSE)

The core concept of this new alloy design philosophy is based on two key principles: (1) designing by microstructure rather than by composition, and (2) utilizing alternative concepts (rather than traditional ones directly linked to the predicted effects of specific elements) as dynamic elements in the desired microstructure. These concepts include configurational entropy of mixing, SFE, lattice misfit, and APBE, among others. The goal is to minimize the reliance on specific raw materials, making alloy development more sustainable. By applying these criteria as design parameters, the inclusion of any element in the final alloy formulation can be validated, avoiding the exclusion of elements typically considered undesirable in e-waste and scrap metals.

One challenge of this approach is the vast number of possible alloy variants when using complex alloys made from waste or scrap, or when combining a wide array of metallic elements. This necessitates the use of powerful computational tools, which should be integrated with data-driven learning systems such as machine learning [150,151]. Fortunately, this approach is well-established in materials science, where modelling (including thermodynamic modelling) combined with artificial intelligence algorithms and machine learning are common practices in MSE[152].

Another challenge is the need for accurate data to support learning-based design systems. This can be addressed through two parallel approaches such as (1) establishing a robust data management system, including the systematic acquisition of external data from open data sources (a practice encouraged by the European Union), and (2) generating large datasets through high-throughput manufacturing and characterization techniques[153]. These approaches are complementary and not mutually exclusive and fully aligned with the so called "autonomous research and development in materials"[154].

Considering these concepts, there are four critical factors when approaching new alloy design: (1) Which raw materials are available? This involves evaluating all metals on the market (including critical or strategic metals) and the potential for using scrap and waste alloys. (2) Which models and optimization methods will be employed (including artificial intelligence)? (3) Which data will be used to support the optimization process (and can high-performance fabrication and characterization generate additional data)? (4) Which design parameters will guide the sought microstructure: high configurational entropy, SFE, lattice misfit, APBE, etc.? It is important to note that these parameters are especially relevant for structural materials; functional materials may require different parameters.

These four factors are directly linked to the microstructure that must be achieved and, consequently, to the material's performance. They are also strongly connected to the processing techniques required to develop the materials. In this new paradigm, we have to consider the revolution that is happening in terms of the new additive manufacturing technologies that also will contribute in the sustainable balance due to its contribution to the use of lesser raw materials, the reduction of weight through better design and a total reduction of the circular $CO_2$ emissions [155].

To shift from focusing on 'appropriate composition' to 'appropriate microstructure,' it is necessary to modify the classical paradigm of MSE. The traditional tetrahedron correlating chemical composition, properties, manufacturing, and performance should be revised. The 'composition' vertex should be replaced by a sub-tetrahedron with four vertices: available feedstocks, design parameters, modelling/optimization, and data generation. This new sub-tetrahedron is linked to processing, which must be optimized in alignment with these new "interstitials" in the tetrahedron as shown in Figure 14. This paradigm shift will enable the use of alternative and more sustainable material sources by focusing not only on composition but also on innovative design and processing parameters that generate cutting-edge microstructures.

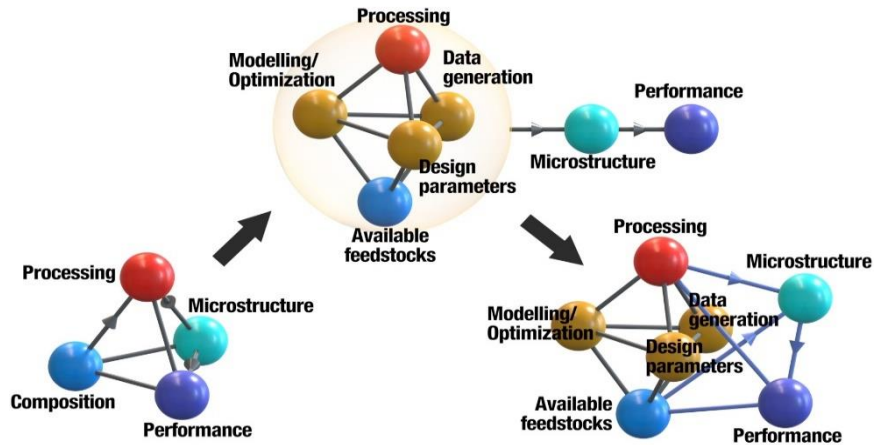

Figure 14. The new tetrahedron for development of HEAs/AHEs.

## 7. Summary

The transition from High-Entropy Alloys (HEAs), also referred to as MPEAs, to AHEs is explored in this discussion. This shift is particularly analysed through the lens of the development of high-entropy steels, superalloys, and intermetallics, highlighting the diversity and performance advantages these alloys can offer. We can emphasize on the following topics:

- Optimization and design of microstructures. In the development of high-performance alloys, there has been a notable focus on optimizing their microstructures. Traditionally, alloy design emphasizes composition and specific elemental interactions. However, in many papers is suggested that alternative criteria can be used to achieve desired properties. Factors like the high entropy of mixing, SFE, lattice misfit, and APBE are being increasingly integrated into the design process. These properties are crucial for enhancing specific performance characteristics, including strength, toughness, and thermal stability.

- Multicomponent alloys from scrap and waste. One of the most innovative aspects of this transition is the shift towards utilizing alloys with a wide array of components. The presence of numerous elements in an alloy without a need to prioritize or discard any specific component has been found to provide substantial benefits. This approach is particularly relevant for utilizing materials sourced from scrap or electronic waste. Such materials offer vast pools of potential alloy compositions, often with millions of combinations, allowing for the creation of alloys that would be difficult to design through traditional methods. This opens up new possibilities for sustainable material sourcing and waste reduction.
- Advanced optimization through artificial intelligence. With such an enormous number of potential combinations in alloy design, there is a growing demand for sophisticated optimization techniques. Traditional trial-and-error methods are no longer sufficient to explore the complexity of these alloys effectively. Artificial intelligence (AI) now plays a critical role in accelerating the alloy design process. AI-driven approaches, particularly machine learning and data-driven modelling, can predict the behaviour of different alloy compositions based on large datasets, ensuring that the most promising formulations are selected. By leveraging AI, it becomes possible to efficiently navigate the vast space of possible alloy combinations, optimizing both the design and performance of these materials.
- Paradigm shift in alloy development. Finally, the integration of these new methodologies necessitates a radical shift in the fundamental principles of alloy development. Traditional alloy design has relied heavily on empirical knowledge and predictable relationships between composition and properties. However, the incorporation of high-entropy concepts, advanced computational modelling, and data-driven strategies requires the adoption of previously unconsidered concepts and strategies. This results in a paradigm shift in Materials Science and Engineering (MSE), as the focus transitions from solely optimizing composition to also considering innovative microstructural design parameters and novel processing techniques. This new approach allows for the development of more sustainable, high-performance materials with a broad range of applications.

This shift from HEAs, also known as MPEAs, to AHEs exemplifies how materials science is evolving, driven by advances in technology, data science, and a more holistic approach to designing alloys for modern applications. Table 2 summarizes the main findings, showcasing the evolution of alloy design from traditional methods to more innovative and sustainable approaches.

Table 2. Key findings for the discussed topics.

| Topic | Key Findings |
|---|---|
| Optimization and design of microstructures | - Traditionally the materials design was focused on optimizing microstructures for high-performance alloys.<br>- Alternative criteria such as high entropy of mixing, stacking fault energy (SFE), lattice misfit, and anti-phase boundary energy (APBE) are now integrated into the design process. |
| Multicomponent alloys from scrap and waste | - Shift towards using alloys with multiple components, without prioritizing or discarding any.<br>- Utilization of scrap and electronic waste opens up millions of possible alloy combinations. |
| | - Provides sustainable material sourcing and waste reduction opportunities. |
| Advanced optimization through artificial intelligence | - AI-driven approaches, particularly machine learning, are crucial for alloy design optimization.<br>- AI helps navigate vast alloy combinations and predict their performance efficiently.<br>- Data-driven modeling ensures selection of the most promising alloy formulations. |
| Paradigm shift in alloy development | - Traditional alloy design based on empirical knowledge is being replaced by new approaches.<br>- High-entropy concepts, computational modeling, and data-driven strategies demand a shift in design principles.<br>- Focus moves from composition optimization to considering microstructural design and novel processing techniques. |
| Evolution of materials science | - The transition from HEAs/MPEAs to AHEs reflects an evolving materials science paradigm.<br>- Advances in technology and data science are driving more holistic alloy design for modern applications, resulting in sustainable and high-performance materials. |

**Declaration of competing interest**

The authors declare that they have no known competing financial interests or personal relationships that could have appeared to influence the work reported in this paper.

**CRediT authorship contribution statement**

José M. Torralba: Conceptualization, Methodology, Supervision, Resources, Writing – original draft; Alberto Meza: Conceptualization, Methodology, Writing – original draft; S. Venkatesh Kumaran: Investigation, Validation, Writing – review and editing; Amir Mostafaei: Visualization, Formal analysis, Writing – review and editing; Ahad Mohammadzadeh: Conceptualization, Investigation, Writing – review and editing

**Acknowledgement**

This investigation was partially supported by the European Union Horizon 2020 research and innovation program (Marie Sklodowska-Curie Individual Fellowships, Grant Agreement 101028155). Amir Mostafaei acknowledges support from the National Science Foundation under grant number CMMI-2339857.